# Enhancement of Critical Current Density

# in low level Al-doped MgB$_2$

A. Berenov[1*], A. Serquis[2,3], X. Z. Liao[2], Y. T. Zhu[2], D.E. Peterson[2], Y. Bugoslavsky[4], K. A. Yates[1], M.G. Blamire[1], L. F. Cohen[4], and J.L. MacManus-Driscoll[1]

[1] Department of Materials Science and Metallurgy, University of Cambridge, Pembroke St., Cambridge, CB2 3QZ, UK
[2] Superconductivity Technology Center, Los Alamos National Laboratory, Los Alamos, NM 87545, USA
[3] Fellowship of CONICET. Permanent address: Centro Atómico Bariloche, Bariloche, Río Negro 8400, Argentina
[4] Department of Physics, Imperial College London, London, SW7 2AZ, UK

Two sets of MgB$_2$ samples doped with up to 5 at. % of Al were prepared in different laboratories using different procedures. Decreases in the 'a' and 'c' lattice parameters were observed with Al doping confirming Al substitution onto the Mg site. The critical temperature ($T_c$) remained largely unchanged with Al doping. For 1 – 2.5 at.% doping, at 20K the in-field critical current densities ($J_c$'s) were enhanced, particularly at lower fields. At 5K, in-field $J_c$ was markedly improved, e.g. at 5T $J_c$ was enhanced by a factor of 20 for a doping level of 1 at.% Al. The improved $J_c$s correlate with increased sample resistivity indicative of an increase in the upper critical field, $H_{c2}$, through alloying.

PACS #: 74.70.Ad, 74.62.Dh, 74.25.Ha

[*] Current address: Department of Materials, Imperial College London, London, SW7 2AZ, UK

MgB$_2$ has been the subject of extensive research since the discovery of its superconductivity in 2001 [1]. A relatively high $T_c$ of 39K, large values of critical current densities, transparency of grain boundary to current flow and simplicity of preparation make MgB$_2$ a promising candidate for low temperature magnetic applications competing with other superconductors like, e.g. Nb$_3$Sn [2]. The idea of improving the field dependent properties of MgB$_2$ through chemical modification has been investigated in various ways. Several studies have shown that nano-scale impurities can be incorporated in MgB$_2$, e.g., Ti [3], Y$_2$O$_3$ [4] and SiC [5,6]. However, the nature and extent of alloying with the MgB$_2$ is still somewhat unclear. So far, the most successful way to alloy and thereby increase $H_{c2}$ has been to react with C from a SiC source or to process in a some background level of oxygen [2]. It appears that only light C substitution of B can be tolerated before $T_c$ drops sharply [7]. Extensive research into substitutional chemistry has shown that only a few elements can substitute onto the Mg site [8] (Al [8-12], Li [13], Cu [14]). However, so far, there have been no reports of improved superconducting properties with doping on the Mg site [11,15,16]. It is possible that the doping levels studied have been too high or that poor microstructures have obscured any improved properties. In the light of recent work by Gurevich *et al*. [2] which suggests the possibility of strong improvements in $H_{c2}$ by controlling intraband scattering rates through selective doping of the Mg and B sites, it is timely to re-visit the issue of doping of the Mg site.

In this paper we report on the effect of low doping levels of Al on the crystal structure and superconducting properties of MgB$_2$. We show for two sets of samples made in different ways and from different purity starting powders and processing methodologies, improvements in the in-field $J_c$ occurs at both 5K and 20K. The improvements coincide with a systematic change in lattice parameters, and an increased sample resistivity, both indicative of alloying. On the other hand, $T_c$ remains high, indicative of light doping.

Two sets of $Mg_{1-x}Al_xB_2$ (x=0, 0.01, 0.025, 0.05) samples were prepared by solid state reaction.

**Set I.** Samples prepared in Cambridge were made from amorphous B powder (Alfa Aesar, 96-98%), Mg powder (Alfa Aesar, 99.6%) and Al powder (Alfa Aesar, 99.97%) using an atomic ratio (Mg,Al):B = 1:2. Pellets were made from the powders and were then covered in powders of identical composition and encapsulated in Ta foil (Advent, 99.9%). Samples were placed in a tubular furnace together with extra Mg rods and synthesised under a flowing 1% $H_2/N_2$ gas mixture with an oxygen partial pressure of around $10^{-10}$ atm. (measured by a YSZ sensor). Samples were reacted at 900°C for 15 min using heating and cooling rates of 15°C/min.

**Set II.** Samples were prepared at Los Alamos from amorphous boron powder (Alfa Aesar, 325 mesh, 99.99%), Mg turnings (Puratronic, 99.98%), and Al powder (Alfa Aesar, 99.5 %) using an atomic ratio of (Mg,Al):B = 1:1. Pellets are made from the powders, and were wrapped in Ta foil with additional Mg turnings, and placed in an alumina crucible inside a tube furnace. The samples were heated at 900°C for one hour under ultra-high purity flowing Ar, cooled down at 0.5°C/min to 500°C, and then furnace cooled to room temperature.

Phase composition and sample crystallinity of set I samples were studied by x-ray diffraction (XRD). Lattice parameters and Mg vacancy concentration were obtained from refinements of XRD data according to the Rietveld method [17]. Particle size and non-uniform strain were estimated from the XRD peak broadening [18]. Sample morphology and chemical composition was studied using scanning electron microscopy (SEM) (set I) and transmission electron microscopy (TEM) (set II).

Magnetisation data were collected using vibrating sample magnetometry (VSM) (set I) and SQUID magnetometry (set II). The temperature dependence of the magnetisation was measured upon heating previously zero-field-cooled samples in an applied field of 1 mT. Magnetisation loops were recorded in magnetic field up to 8 T, with prior excursion to the reverse-polarity field of 1 or 3T. For set I, the characteristic length scale of the connected current-carrying regions (at T=20 K and in fields of up to 3T) were estimated according to the method described by Angadi *et al*. [19] and were found to be close to the sample dimensions. Hence, $J_c$s were calculated using the Bean critical state model [20] using full sample dimensions. Resistivity was measured by the 4-point Van Der Pauw technique (set I).

XRD showed the formation of the hexagonal $MgB_2$ phase. Mg and MgO impurities were the only secondary phases detected in the samples. About 3 wt.%. of MgO was observed in all the samples. No Al-rich phases were detected. The effect of Al doping on the lattice parameters is shown in Figure 1. The observed values of the lattice parameters compare reasonably well with the literature data [10,11] except for the data of Toulemonde *et al.* [9] where samples were prepared under a high pressure of 3.5 GPa. A linear decrease of both 'a' and 'c' lattice parameters (larger than the respective standard deviations), and the absence of Al-rich phases confirm Al substitution into $MgB_2$. Smaller changes in the 'a' lattice parameter (-4.8 x $10^{-4}$ Å/at. %) as compared to the 'c' lattice parameter (-2.6 x $10^{-3}$ Å /at. %) correlate with the rigidity of the B-B bonds in the boron layers in the $MgB_2$ structure.

The values of non-uniform strain as calculated from the XRD data for set I are shown in Table 1. The calculated crystallite size was in the range 70 – 130 nm. No data for crystallite size and non-uniform strain is available for the 5% doped sample due to the lack of linearity observed in the Williamson-Hall plot which was used to measure these parameters. It is possible that peak broadening in the XRD plot of the 5% doped sample was partially caused by the coexistence of two $Mg_{1-x}Al_xB_2$ phases with close lattice parameters as previously reported for 10% Al doped $MgB_2$ [8]. The increase of non-uniform strain with doping was caused by a substitution of the smaller Al ion (67.5 pm) onto the Mg site (86.0 pm [21]) and/or formation of Mg vacancies during Al doping. Serquis *et al.* [22] showed a linear relationship between Mg occupancy and non-uniform strain in the undoped $MgB_2$.

Extra positive charge results from substitution of $Al^{3+}$ on $Mg^{2+}$ sites can be compensated either by formation of Mg vacancies and/or B interstitial as shown by the following equations, using Kroger-Vink notation:

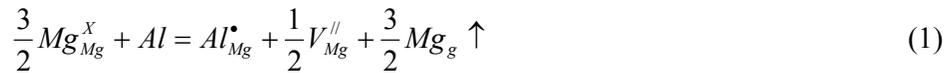

$$\frac{3}{2}Mg_{Mg}^{X} + Al = Al_{Mg}^{\bullet} + \frac{1}{2}V_{Mg}^{//} + \frac{3}{2}Mg_{g}\uparrow \qquad (1)$$

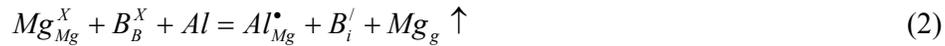

$$Mg_{Mg}^{X} + B_{B}^{X} + Al = Al_{Mg}^{\bullet} + B_{i}^{/} + Mg_{g}\uparrow \qquad (2)$$

The formation of B interstitials has been suggested by Zhao *et al.* [23]. However, the formation of Mg vacancies is more likely to occur due to a high volatility of Mg [24]. Indeed, improvement of the Rietveld refinements for set I was achieved when Mg vacancies were introduced in the refinements. Mg vacancies were present in the undoped sample to a maxiumum level of ~ 7% and that the extent of Mg non-stoichiometry increased with Al doping, up to a maximum level of ~9% for 5% substitution. Hence, equation 1 most closely describes the defect equilibria in $Mg_{1-x}Al_xB_2$ although it does not include the presence of vacancies in the undoped sample. We note that vacancy concentration values are rather high but not implausible, because, as with the Al doping, they can arise from a charge compensation effect associated with impurity doping from the impure starting powders.

Set I samples consisted of well connected micron size crystallites as observed in SEM (not shown). A slight increase of grain size was observed with Al doping (from ~

0.6 μm to ~ 1 μm for the 0% and 5% doped samples, respectively). TEM studies of set II samples showed the presence of nano-particles of MgO at the grain boundaries (dark regions) in the 2.5% Al doped sample (Figure 2a) but much less for lower doping levels, as shown for example for the 1% doped sample (Figure 2b). The MgO most likely arises from oxidation of Mg, rejected from $MgB_2$ grains to grain boundaries, after occupation of the Mg site by Al, according to equation 1. The defect density (mostly dislocations) increased upon Al doping which is consistent with the increased strain in the lattice. Some segregation of Al to the grain boundaries was also observed in the most highly doped (5%) sample.

The onset of the $T_c$ transition was 37.4 – 39.0 K and only slightly decreased with doping (Table 1). The transition width was around 8 – 9 K for samples prepared from impure B (Set I) and around 1 K for the samples prepared from pure B (Set II). Although small particle sizes were found in the $MgB_2$ samples (set I), it is unlikely that proximity effects [25] resulted in the increased width of superconducting transition since the length scale measurements implied full connectivity within the samples. Hence, it is mostly likely that the broadened transition for set I arose from the presence of impurities.

The field dependences of critical current densities for sets I and II samples are shown in Figure 3. At 20K, typical $J_c$s (around 3-4x$10^5$ A/cm$^2$) were measured for both sets of undoped samples. $J_c$ was enhanced in fields of up to at least 4T with doping levels of 2.5 % (I) and 1 % Al (II). Considering that the two sets of samples were prepared in quite different ways, it is not surprising that the optimal *nominal* doping level was different between sample sets.

Only set II was measured at 5K. Compared to the 20K data, for doping levels of up to 2.5 %, even larger enhancements of $J_c$ were observed, and this occurred over the whole range of magnetic fields applied (to 7T), e.g. $J_c$ was increased by an order of magnitude at 5T for the 1at. % doped sample.

The improvement of $J_c$ performance with Al doping is far less than in B-site doped samples [6]. Nevertheless, our results compare favourably with previous studies of low level Al doped (~5 at.%) samples, where the doping caused sharp reductions of $J_c$ over the temperature range 5 – 35 K [9,15]. Most importantly, the possibility now exists to tune the intraband scattering rates by selectively doping *both* the B and Mg sites.

Owing to intergranular effects, resistivity is not always a direct measure of intragranular scattering [26]. Nevertheless, we find a good correlation between room temperature resistivity (Table 1) and self-field $J_c$ at 20K (Figure 4) indicative of an increased $H_{c2}$ through alloying [6]. We observe an increase in resistivity up to 2.5 %, as would be expected, but then a decrease for the 5% sample.

The low resistivity for the 5% sample is rather surprising particularly in view of both the increased alloying as observed from the x-ray data and the increased level of non-conductive nano-MgO observed at the grain boundaries with doping. However, as previously mentioned in discussion of the x-ray data, it is possible that decomposition into two $Mg_{1-x}Al_xB_2$ phases occurs.

Improved performance through alloying rather than changes to the pinning mechanism was corroborated from curves of the reduced flux pinning force ($F_P/F_{Pmax}$ where $F_p = J_c B$) as a function of the reduced magnetic field ($B/B_{irr}$). At 20K, for set I samples the curves overlapped (not shown) implying an identical pinning mechanism [15].

The reason for the much greater improved field performance for the 5K data compared to the 20K data is most likely because of the presence of the nano-MgO at the grain boundaries which is more deleterious at the higher measurement temperature. It is likely that processing refinements (e.g. re-grinding and re-sintering of samples) could clean the grain boundaries and, therefore, lead to further improvements of the in-field performance.

In summary, polycrystalline $MgB_2$ samples with up to 5 at.% Al doping of the Mg site were synthesized in two different laboratories using different purity starting chemicals and reaction procedures. At 20K, small amounts of Al doping (less than 2.5 at.%) enhance the self-field $J_c$ values by a factor of ~2 and also gave improved field performance up to at least 4T. At 5K, the in-field $J_c$ was markedly improved, especially at higher fields. $T_c$s remained largely unchanged upon Al doping. The possibility now exists to selectively dope both the Mg and B sites to yield further improvements in $H_{c2}$.

| Nominal doping (at. %) | Non-uniform Strain (%) | $T_c$ onset (K) | $\Delta T_c$ (K) | $J_c$ (A cm$^{-2}$) at 5K, 0T | $J_c$ (A cm$^{-2}$) at 20K, 0T | Resistivity ($\mu\Omega$ cm) 300K |
|---|---|---|---|---|---|---|
| Set I | | | | | | |
| 0 | 0.38 | 38.4 | NA | NA | 2.70x10$^5$ | 41.6 |
| 1 | 0.43 | 39.0 | 9.4 | NA | 3.67x10$^5$ | 120.3 |
| 2.5 | 0.44 | 38.6 | 8.3 | NA | 4.47x10$^5$ | 106.5 |
| 5 | NA | 37.4 | 9.1 | NA | 2.80x10$^5$ | 79.0 |
| Set II | | | | | | |
| 0 | NA | 39.2 | 0.8 | 3.87x10$^5$ | 4.20x10$^5$ | NA |
| 1 | NA | 38.5 | 1 | 6.00x10$^5$ | 4.15x10$^5$ | NA |
| 2.5 | NA | 38.5 | 1 | 4.78x10$^5$ | 3.17x10$^5$ | NA |
| 5 | NA | 38 | 1.4 | 2.70x10$^5$ | 1.50x10$^5$ | NA |

Table 1: Non-uniform strain and superconducting parameters of Al-doped MgB$_2$.

**Figure Captions**

Figure 1. The effect of Al doping on (a) 'a' and (b) c' lattice parameters in MgB$_2$ (Set I).

Figure 2. TEM images of 2.5% (a) and 1% (b) Al doped MgB$_2$ sample (Set II). Very dark areas (some are indicated using white arrows) at grain boundaries in (a) are nano-sized MgO.

Figure 3. Critical current densities ($J_c$s) of MgB$_2$ samples (a) set I, at 20K, (b) set II at 20K and (c) set II at 5K

Figure 4. Correlation between $J_c$ at 20 K, 0 T and room temperature resistance for set I samples.

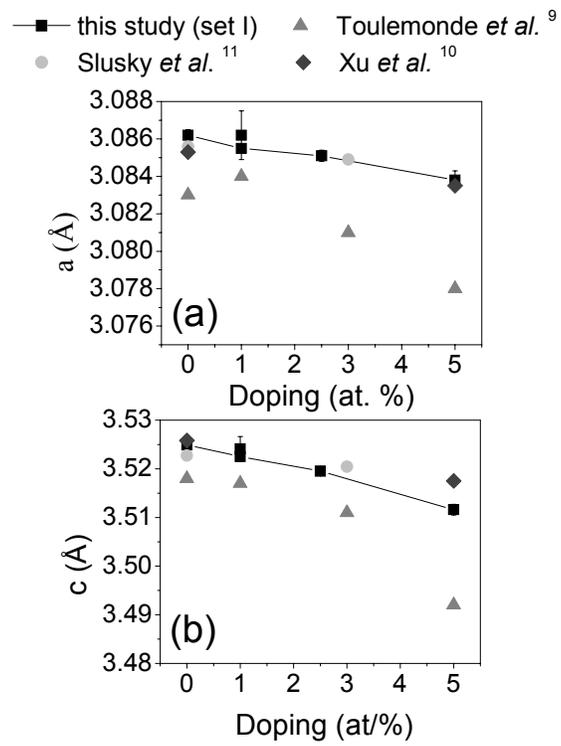

Figure 1.

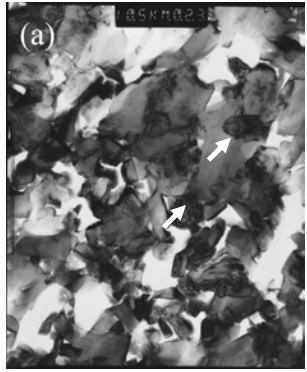
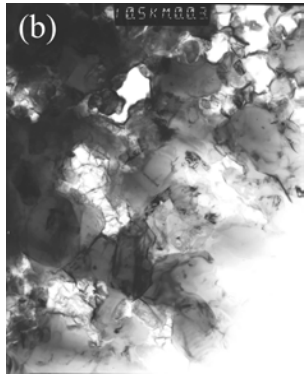

Figure 2.

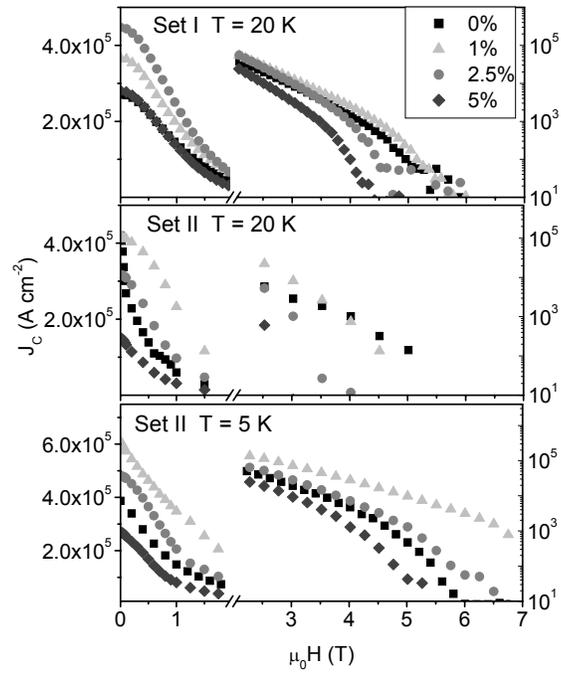

Figure 3.

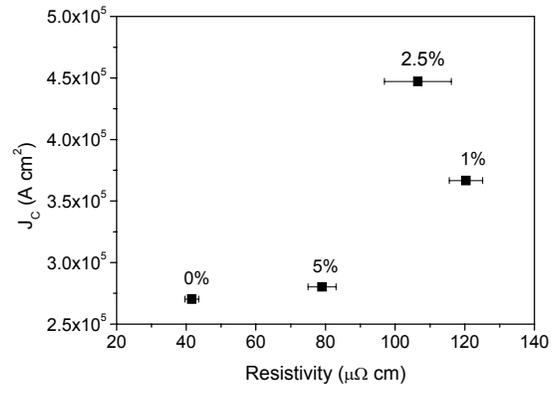

Figure 4.